\documentclass[twocolumn,times,tighten]{aastex701}

\usepackage{xspace}
\usepackage[T1]{fontenc}
\usepackage{lmodern}

\newcommand{\FeKa}{Fe K\ensuremath{\alpha}\xspace}
\newcommand{\FeKb}{Fe K\ensuremath{\beta}\xspace}
\newcommand{\civ}{\ion{C}{4}\xspace}
\newcommand{\fexxv}{\ion{Fe}{25}\xspace}
\newcommand{\fexxvi}{\ion{Fe}{26}\xspace}
\newcommand{\kms}{\ensuremath{\mathrm{km\ s^{-1}}}\xspace}
\newcommand{\NH}{\ensuremath{N_{\mathrm{H}}}\xspace}
\newcommand{\vout}{\ensuremath{v_{\mathrm{out}}}\xspace}
\newcommand{\sigv}{\ensuremath{\sigma_{v}}\xspace}
\newcommand{\cf}{\ensuremath{C_f}\xspace}

\newcommand{\xabs}{\xspace{\tt xabs}\xspace}
\newcommand{\delt}{\xspace{\tt delt}\xspace}
\newcommand{\vgau}{\xspace{\tt vgau}\xspace}
\newcommand{\pow}{\xspace{\tt pow}\xspace}
\newcommand{\comt}{\xspace{\tt comt}\xspace}

\newcommand{\nustar}{{\it NuSTAR}\xspace}
\newcommand{\xmm}{{\it XMM-Newton}\xspace}
\newcommand{\xrism}{{XRISM}\xspace}
\newcommand{\chandra}{{\it Chandra}\xspace}
\newcommand{\swift}{{\it Swift}\xspace}

\newcommand{\ergflux}{{\ensuremath{\rm{erg\ cm}^{-2}\ \rm{s}^{-1}}}\xspace}

\newcommand{\cm}{{\ensuremath{\rm{cm}^{-2}}}\xspace}
\newcommand{\spex}{\xspace{\tt SPEX}\xspace}
\newcommand{\pion}{\xspace{\tt pion}\xspace}

\newcommand{\logxi}{\ensuremath{{\log \xi}}\xspace}

\newcommand{\lya}{Ly\ensuremath{\alpha}\xspace}

\newcommand{\ngc}{{NGC~5548}\xspace}

\usepackage{xcolor}

\usepackage{xfrac}
\shorttitle{Dissecting the Nuclear Structure of NGC 5548 with XRISM. I.}
\shortauthors{Mehdipour et al.}
\received{April 10, 2026}
\revised{June 30, 2026}
\accepted{August 4, 2026}
\graphicspath{{./}{figures/}}
\begin{document}

\title{\large Dissecting the Nuclear Structure of NGC 5548 with XRISM \\ [1ex] \normalsize \textit{I. Physical Properties of the Highly Ionized Outflows}}

\author[0000-0002-4992-4664,gname=Missagh,sname=Mehdipour]{Missagh Mehdipour}
\affiliation{Department of Astronomy, University of Michigan, 1085 South University Avenue, Ann Arbor, MI, 48109, USA}
\email[show]{missagh@umich.edu}

\author[0000-0003-2869-7682,gname=Jon,sname=Miller]{Jon M. Miller}
\affiliation{Department of Astronomy, University of Michigan, 1085 South University Avenue, Ann Arbor, MI, 48109, USA}
\email{jonmm@umich.edu}

\author[0000-0001-5540-2822,gname=Jelle,sname=Kaastra]{Jelle S. Kaastra}
\affiliation{SRON Netherlands Institute for Space Research, Niels Bohrweg 4, 2333 CA Leiden, the Netherlands}
\affiliation{Leiden Observatory, Leiden University, PO Box 9513, 2300 RA Leiden, the Netherlands}
\email{j.s.kaastra@sron.nl}

\author[0000-0002-2180-8266,gname=Gerard,sname=Kriss]{Gerard A. Kriss}
\affiliation{Space Telescope Science Institute, 3700 San Martin Drive, Baltimore, MD 21218, USA}
\email{gak@stsci.edu}

\author[0000-0001-5709-7606,gname=Keigo,sname=Fukumura]{Keigo Fukumura}
\affiliation{Department of Physics and Astronomy, James Madison University, Harrisonburg, VA 22807, USA}
\email{fukumukx@jmu.edu}

\author[0000-0001-9911-7038,gname=Liyi,sname=Gu]{Liyi Gu}
\affiliation{SRON Netherlands Institute for Space Research, Niels Bohrweg 4, 2333 CA Leiden, the Netherlands}
\affiliation{Leiden Observatory, Leiden University, PO Box 9513, 2300 RA Leiden, the Netherlands}
\email{L.Gu@sron.nl}

\author[0000-0002-3687-6552,gname=Doyee,sname=Byun]{Doyee Byun}
\affiliation{Department of Astronomy, University of Michigan, 1085 South University Avenue, Ann Arbor, MI, 48109, USA}
\email{doyeeb@umich.edu}

\author[0000-0002-7129-4654,gname=Xin,sname=Xiang]{Xin Xiang}
\affiliation{Department of Astronomy, University of Michigan, 1085 South University Avenue, Ann Arbor, MI, 48109, USA}
\email{xinxiang@umich.edu}

\author[0000-0001-9735-4873,gname=Ehud,sname=Behar]{Ehud Behar}
\affiliation{Department of Physics, Technion, Haifa 32000, Israel}
\email{behar@physics.technion.ac.il}

\author[0000-0003-2663-1954,gname=Laura,sname=Brenneman]{Laura W. Brenneman}
\affiliation{Center for Astrophysics -- Harvard \& Smithsonian, 60 Garden Street, Cambridge, MA 02138, USA}
\email{lbrenneman@cfa.harvard.edu}

\author[0000-0001-8470-749X,gname=Elisa,sname=Costantini]{Elisa Costantini}
\affiliation{SRON Netherlands Institute for Space Research, Niels Bohrweg 4, 2333 CA Leiden, the Netherlands}
\affiliation{Anton Pannekoek Institute, University of Amsterdam, Postbus 94249, 1090 GE Amsterdam, The Netherlands}
\email{E.Costantini@sron.nl}

\author[0000-0002-0964-7500,gname=Maryam,sname=Dehghanian]{Maryam Dehghanian}
\affiliation{Department of Physics and Astronomy, University of Kentucky, Lexington, KY 40506, USA}
\email{m.dehghanian@uky.edu}

\author[0000-0001-5924-8818,gname=Jacobo,sname=Ebrero]{Jacobo Ebrero}
\affiliation{Telespazio UK for the European Space Agency (ESA), European Space Astronomy Centre (ESAC), Camino Bajo del Castillo, s/n, 28692 Villanueva de la Ca\~{n}ada, Madrid, Spain}
\email{jacobo.ebrero.carrero@ext.esa.int}

\author[0000-0003-2754-9258,gname=Massimo,sname=Gaspari]{Massimo Gaspari}
\affiliation{Department of Physics, Informatics and Mathematics, University of Modena and Reggio Emilia, 41125 Modena, Italy}
\email{massimo.gaspari@unimore.it}

\author[0000-0002-7292-6852,gname=Anna,sname=Juráňová]{Anna Juráňová}
\affiliation{MIT Kavli Institute for Astrophysics and Space Research, Massachusetts Institute of Technology, Cambridge, MA 02139, USA}
\email{ajuran@mit.edu}

\author[0000-0003-0172-0854,gname=Erin,sname=Kara]{Erin Kara}
\affiliation{MIT Kavli Institute for Astrophysics and Space Research, Massachusetts Institute of Technology, Cambridge, MA 02139, USA}
\email{ekara@mit.edu}

\author[0000-0001-5493-7585,gname=Chen,sname=Li]{Chen Li}
\affiliation{Leiden Observatory, Leiden University, PO Box 9513, 2300 RA Leiden, the Netherlands}
\affiliation{SRON Netherlands Institute for Space Research, Niels Bohrweg 4, 2333 CA Leiden, the Netherlands}
\email{c.li@sron.nl}

\author[0000-0001-7557-9713,gname=Junjie,sname=Mao]{Junjie Mao}
\affiliation{Department of Astronomy, Tsinghua University, Haidian DS 100084, Beijing, People’s Republic of China}
\email{jmao@tsinghua.edu.cn}

\author[0000-0001-6020-517X,gname=Hirofumi,sname=Noda]{Hirofumi Noda}
\affiliation{Astronomical Institute, Tohoku University, 6-3 Aramakiazaaoba, Aoba-ku, Sendai, Miyagi 980-8578, Japan}
\email{hirofumi.noda@astr.tohoku.ac.jp}

\author[0000-0003-4504-2557,gname=Anna,sname=Ogorzalek]{Anna Ogorzalek}
\affiliation{Department of Astronomy, University of Maryland, College Park, MD 20742, USA}
\affiliation{NASA / Goddard Space Flight Center, Greenbelt, MD 20771, USA}
\affiliation{Center for Research and Exploration in Space Science and Technology, NASA / GSFC (CRESST II), Greenbelt, MD 20771, USA}
\email{ogoann@umd.edu}

\author[0000-0002-1049-3182,gname=Ioanna,sname=Psaradaki]{Ioanna Psaradaki}
\affiliation{European Space Agency, European Space Research and Technology Center, Keplerlaan 1, 2201 AZ Noordwijk, The Netherlands}
\email{Ioanna.Psaradaki@esa.int}

\author[0000-0002-5359-9497,gname=Daniele,sname=Rogantini]{Daniele Rogantini}
\affiliation{Department of Astronomy and Astrophysics, University of Chicago, Chicago, IL 60637, USA}
\affiliation{MIT Kavli Institute for Astrophysics and Space Research, Massachusetts Institute of Technology, Cambridge, MA 02139, USA}
\email{danieler@uchicago.edu}

\author[0000-0002-8163-8852,gname=Sascha,sname=Zeegers]{Sascha T. Zeegers}
\affiliation{SRON Netherlands Institute for Space Research, Niels Bohrweg 4, 2333 CA Leiden, the Netherlands}
\affiliation{Anton Pannekoek Institute, University of Amsterdam, Postbus 94249, 1090 GE Amsterdam, The Netherlands}
\email{S.T.Zeegers@sron.nl}

\author[0000-0003-1252-8227,gname=Keqin,sname=Zhao]{Keqin Zhao}
\affiliation{Leiden Observatory, Leiden University, PO Box 9513, 2300 RA Leiden, the Netherlands}
\affiliation{SRON Netherlands Institute for Space Research, Niels Bohrweg 4, 2333 CA Leiden, the Netherlands}
\email{kzhao@strw.leidenuniv.nl}

\begin{abstract}
We present a detailed spectral analysis of an X-Ray Imaging and Spectroscopy Mission (\xrism) observation of the prototypical Seyfert 1 galaxy \ngc.  
\xrism’s Resolve microcalorimeter reveals, for the first time, highly ionized outflows in this active galactic nucleus (AGN) through the detection of \fexxv and \fexxvi absorption lines in the Fe~K band.  
Modeling the \xrism/Resolve spectrum alongside \xmm Reflection Grating Spectrometer (RGS) data allows us to probe the ionization and kinematic structure of the outflows in this AGN.  
We identify four distinct ionization components, with ionization parameters \logxi ranging from 0.9 to 3.4.  
Three of these components are further resolved into two velocity sub-components, demonstrating the multiphase structure of the outflows.  
The measured outflow velocities span 240 to 2730~\kms.  
We find a trend of increasing column density with ionization parameter ($\xi$), along with a general pattern of increasing outflow velocity with $\xi$.  
The \xrism/Resolve spectrum provides a far more detailed absorption measure distribution (AMD) than was previously possible, revealing two distinct slopes above and below ${\log\xi \sim 2.6}$.  
A comparison of the \fexxv absorption line profile with UV absorption lines (\civ and \lya) observed with the Hubble Space Telescope reveals both overlaps and deviations.
The \xrism/Resolve results suggest a multiphase, clumpy outflow in \ngc, consistent with a ``hybrid wind'' scenario in which the observed parameter trends arise from multiple origins and driving mechanisms.
\end{abstract}
\keywords{\uat{X-ray active galactic nuclei}{2035} --- \uat{Seyfert galaxies}{1447} --- \uat{High resolution spectroscopy}{2096} ---  \uat{Photoionization}{2060} --- \uat{Galaxy winds}{626}}
\section{Introduction} 
\label{sect_intro}
Winds from active galactic nuclei (AGNs) provide a vital link between supermassive black holes (SMBHs) and their host galaxies \citep{King15,Gasp17,Harr18}. Observed correlations such as the M-$\sigma$ relation \citep{Korm13} suggest that SMBHs and their hosts co-evolve through feedback processes, in which AGN winds likely play a central role. Yet the dynamics, kinematics, and ionization structure of outflows, from the accretion disk to kiloparsec scales, remain poorly constrained, limiting our understanding of how efficiently they transport momentum and energy. Ionized outflows are observed across multiple spatial scales \citep{Laha21,Gall23}, but their physical connection and origin remain unclear. Fundamental questions persist regarding their driving mechanisms (thermal, radiative, or magnetic) and whether their large-scale impact is primarily momentum- or energy-driven.
Beyond the driving mechanism, an equally important open question is whether the absorbers detected at different ionizations trace distinct outflows or, instead, different thermodynamic phases of a single structured flow. In this broader picture, the hot X-ray absorbing phase may coexist with cooler and denser material through ongoing condensation, entrainment, mixing, and reheating, producing a clumpy, radially stratified flow rather than a single smooth wind. Establishing whether AGN outflows are organized in this multiphase way is essential for connecting the innermost launching region to the larger-scale feedback cycle (for a review, see \citealt{Gasp20}).

The X-ray Imaging and Spectroscopy Mission (\xrism), equipped with its Resolve microcalorimeter \citep{Tash20,Tash25}, provides an unprecedented combination of high energy resolution (4.5 eV full width at half maximum, FWHM) and sensitivity in the 1.8–12 keV band, making it a powerful tool for investigating highly ionized outflows and ultra-fast outflows (UFOs; \citealt{Tomb10}) that manifest as X-ray absorption features in the Fe K band (6–9 keV). These outflows, which are faster and more energetic than classical ``warm absorbers'' \citep{Blu05}, uniquely probe the innermost regions of AGN outflows, making them powerful diagnostics of the physical conditions near the SMBH.
\xrism/Resolve's non-dispersive, high-resolution spectroscopy facilitates the detection of faint and narrow Fe K absorption lines, enabling the identification of multiple velocity components and small velocity shifts in highly-ionized absorption lines, such as \fexxv and \fexxvi lines.
This capability enables the separation of complex absorption and emission features associated with different types of outflows, allowing precise measurements of outflow kinematics and ionization states.
Thus, \xrism\ measurements are critical for constraining the launching radius and the energetics of the fastest components of AGN outflows, and they have so far shed important new light on outflows in AGN.

Recent \xrism/Resolve observations have revealed a remarkable level of complexity in AGN outflows. In the luminous quasar PDS~456, Resolve resolves multiple UFO components and a highly stratified ionization structure \citep{XRISM25,Xu25}, while in the Seyfert galaxies NGC~4151 \citep{Xian25,xrism24} and NGC~3783 \citep{Mehd25,Li26}, it detects multi-phase winds spanning slow warm absorbers to sub-relativistic UFOs, revealing substantial ionization and kinematic substructure. In NGC~3783, \xrism\ also captured the emergence of a UFO during the decay of a soft X-ray/UV flare \citep{Gu25}. Similar wind studies have now been reported for NGC~3516 \citep{Jura25}, NGC~4051 \citep{Reev26}, Mrk~279 \citep{Mill25}, PG~1211+143 \citep{Mizu26}, NGC~1365 \citep{Zaid26}, MCG--6-30-15 \citep{Bren25}, IRAS~05189--2524 \citep{Noda25b}, and Centaurus~A \citep{Kaya26}. Collectively, these results demonstrate the power of \xrism\ to resolve the ionization and velocity structure of AGN winds and suggest that such outflows are often highly structured and multiphase.

The archetypal Seyfert 1 galaxy \ngc has been at the forefront of discoveries on AGN outflows, owing to large campaigns and extensive multi-wavelength observations \citep{Kaas14,DeRo15}. 
\ngc is one of the best-studied low $L/L_{\rm Edd}$ AGN (${\sim 0.03}$, \citealt{Mehd24}), and thus understanding its winds provides insight into an AGN population that eROSITA finds to be the most numerous in the local universe \citep{Liu22}.
The link between X-ray obscuration and wind activity was first established during the 2013 campaign of \ngc \citep{Kaas14,Arav14,Meh15a,Mao17}. At that time, the source became obscured in X-rays, while new broad and blueshifted UV absorption lines, such as \civ, appeared in the HST/COS spectra \citep{Kris19b,Mehd22c}. 
Unlike the commonly observed warm absorber outflows at pc-scale distances, this obscuration is caused by a massive stream of outflowing gas in the vicinity of the accretion disk, extending to and beyond the broad-line region (BLR) \citep{Kaas14,Mehd22c,Fuku24}. This component is commonly referred to as an obscuring wind, or simply the obscurer. 
Interestingly, while the inner obscurer shields outflows at larger radii that are observed in the UV band with HST/COS \citep{Kris19b,Mehd22c}, thus lowering their ionization, the ionized outflows detected in the X-ray band with \chandra/HETG \citep{Mehd24} remain unaffected by the obscurer.

Long-term \swift\ and HST monitoring of \ngc \citep{Mehd22c} shows that the heavy obscuration has gradually declined over the last decade.
The HST/COS data suggest that the obscurer shields the surrounding gas, thereby influencing the appearance of both the warm absorbers \citep{Dehg19,Mehd22c} and the BLR emission lines \citep{Dehg19b}.
\chandra/HETG spectroscopy of \ngc \citep{Mehd24} reveals the presence of highly ionized \ion{Si}{14} absorption; however, the full extent of the highly ionized outflows cannot be probed with HETG due to the limited signal-to-noise in the Fe K band \citep{Mehd24}. As a result, their properties have remained poorly constrained until now.
The \xrism/Resolve observation of \ngc in Cycle~1 provides the first clear detection of highly ionized outflows in the Fe~K band in this AGN.
This new \xrism\ view of \ngc provides a valuable opportunity to test whether the highly ionized absorber is simply an additional zone or, instead, represents the hot phase of a broader multiphase structure linking hot, warm, and cooler gas across a range of radii (e.g., \citealt{Gasp17,McKi22}).

In this paper, we present the first analysis of the \xrism/Resolve observation of \ngc, which accumulated a total Resolve exposure of 301 ks. Our primary focus is the joint modeling of the time-averaged \xrism/Resolve spectrum with \xmm/RGS data, along with a comparison of the outflow kinematics with HST/COS observations. This multi-instrument approach allows us to investigate the ionization and kinematic structure of the outflows in \ngc. Other aspects, including reflection and spectral variability, will be presented in subsequent papers in this series.

\section{Observations and Data Reduction} 
\label{sect_data}

The \xrism\ Cycle~1 observation of \ngc\ (Obs. ID: 201081010) was carried out from July~2 to July~9, 2025.
We also analyze new \xmm spectra (176 ks; Obs. IDs: 0943020101 and 0943020201), which were obtained jointly with \nustar (100 ks, Obs. IDs: 61001003002 and 61001003004) and HST/COS (2 orbits, Program ID: 17591) data on February 1-2, 2025. We describe the coupling of our Resolve and RGS spectral analyses in Sect. \ref{sect_model} (see Figs. \ref{fig_full} and \ref{fig_spec}).
The observed X-ray flux at the time of the \xrism observation is ${F_{2\text{--}10\,{\rm keV}} = 3.6 \times 10^{-11}}$~\ergflux, compared to ${F_{2\text{--}10\,{\rm keV}} = 2.7 \times 10^{-11}}$~\ergflux during the \xmm observation.
These values are generally consistent with the average flux from \swift monitoring in recent years \citep{Mehd22c}.

\subsection{XRISM Data}
\label{sect_resolve}

Our \xrism\ observation of \ngc\ was carried out with the gate valve closed and the Resolve filter wheel in the open configuration. 
We used {\tt HEASoft v6.35.2} for the reduction and preparation of the Resolve data.
Data analysis was performed using the \xrism\ {\tt CALDB version 12} (20250915 release) and the {\tt ftools} package, with additional screening and energy-dependent rise-time cuts applied following the XRISM Quick Start Guide Version~3.2. 
The used {\tt CALDB} includes the latest updates to the Resolve energy scale and line-spread function, based on in-orbit calibration data. 
After applying good time interval (GTI) filtering, the total effective exposure time is 301 ks.

We selected only the high-resolution primary (Hp) events for our analysis \citep{Ishi18}, as nearly all detected events are expected to fall into this category.
To avoid contamination from ``pseudo'' low-resolution secondary (Ls) events, and to prevent the associated normalization errors these introduce in the response matrix file (RMF), all Ls events were removed prior to and contemporaneously with RMF generation \citep{TTWOF}. The RMF was then produced using the extra-large (``{\tt X}'') option, which incorporates all known instrumental effects.
Gain tracking was carried out using 24 fiducial measurements of the Mn K$\alpha$ line from the $^{55}$Fe calibration source in the filter wheel. Among the 36 pixels, pixel~12 (the dedicated calibration pixel) and pixel~27 (which shows irregular gain jumps) were excluded from the analysis.
We accounted for the contribution of the Non-X-ray Background (NXB) by running the {\tt rslnxbgen} tool to extract the NXB spectrum from the entire array (except pixels~12 and~27).
We then applied the NXB spectral model template developed by the XRISM team to fit the normalization of the continuum and the emission features, following the instructions provided on the Non-X-ray Background Database and Tools webpage.

\subsection{XMM-Newton Data}
\label{sect_rgs}

The \xmm data were reduced using the Science Analysis System (SAS v22.1.0). The RGS \citep{denH01} instruments were operated in spectroscopy mode, and the data were processed with the {\tt rgsproc} pipeline, which extracts the source and background spectra and generates the corresponding response matrices. Time intervals with background count rates exceeding $0.1~\mathrm{count\ s^{-1}}$ in CCD~9 were filtered out. The first-order spectra from RGS1 and RGS2 were then combined with the {\tt rgscombine} task to produce a single RGS spectrum for use in the spectral analysis.

The OM images were obtained with the {\it V}, {\it B}, {\it U}, {\it UVW1}, {\it UVM2}, and {\it UVW2} filters. Raw images were processed using the {\tt omichain} pipeline, which applies all standard corrections, including the removal of Modulo-8 fixed-pattern noise, to produce calibrated data products suitable for photometric analysis. Source and background regions were defined to extract count rates, and the resulting measurements were corrected for the point-spread function (PSF), coincidence losses, and the time-dependent sensitivity of the detector. Aperture photometry was carried out using a 12''-diameter circular aperture, consistent with calibration recommendations. The background was estimated from a nearby, source-free aperture of the same size.

\subsection{HST Data}
\label{sect_cos}

Our new HST/COS observations were obtained with the G130M and G160M gratings to cover the \lya and \civ spectral lines, respectively. The data were processed using the latest calibration pipeline, {\tt CalCOS v3.6.1}, and the wavelength solution was verified by cross-checking the positions of known Galactic interstellar medium lines. All individual COS exposures were then co-added into a single calibrated spectrum and binned by four pixels to improve the signal-to-noise ratio (S/N). This binning approach still oversamples the $\sim$10-pixel resolution element of the far-UV detector \citep{Fox18}. Additional details regarding COS data preparation for \ngc are provided in previous HST/COS analyses \citep{Kris19b,Mehd22c}.

\section{Spectroscopy and Modeling} 
\label{sect_model}
\begin{figure}[!t]
\centering
\resizebox{\hsize}{!}{\includegraphics[angle=0]{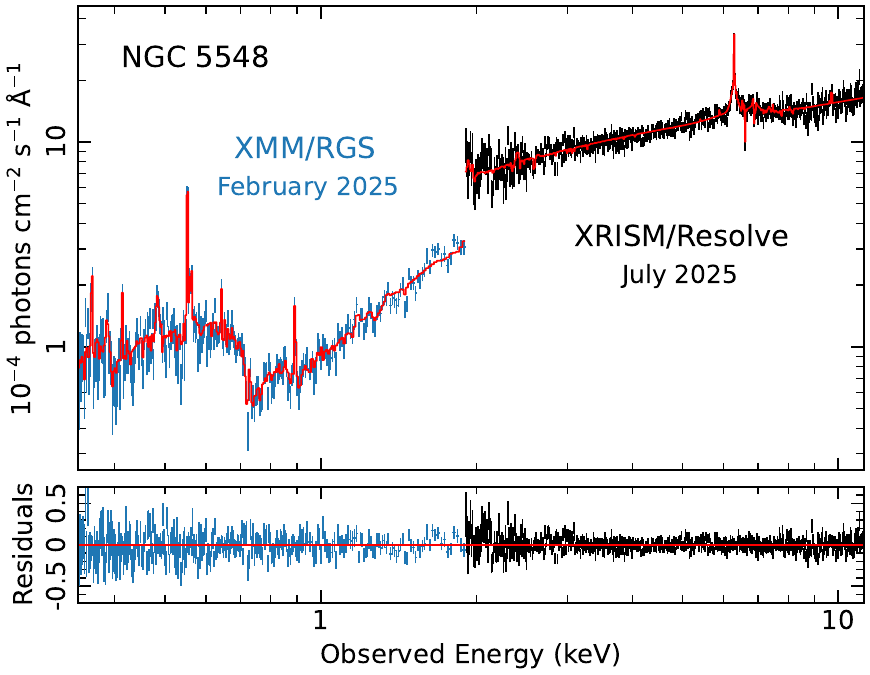}}
\vspace{-0.4cm}
\caption{Overview of the \xrism/Resolve and \xmm/RGS spectra of \ngc, along with our best-fit model. For clarity, the spectra have been further binned. The best-fit model (Tables~\ref{table_cont} and \ref{table_para}), fitted across the full spectral ranges of Resolve and RGS, is shown in red. Residuals are plotted as (data $-$ model) / model.}
\label{fig_full}
\end{figure}

\begin{figure*}[!t]
\centering
\resizebox{0.91\hsize}{!}{\includegraphics[angle=0]{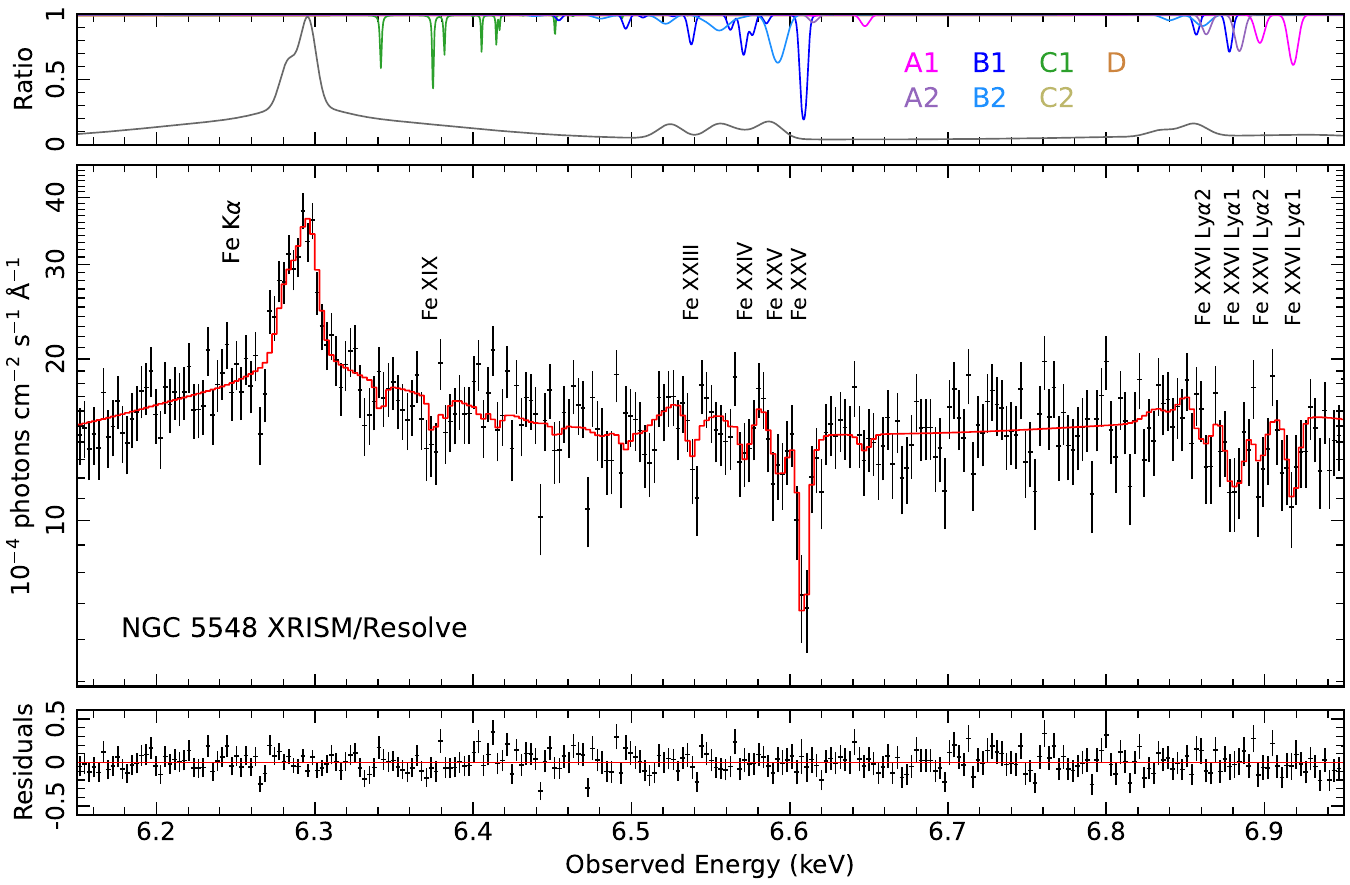}}
\resizebox{0.91\hsize}{!}{\includegraphics[angle=0]{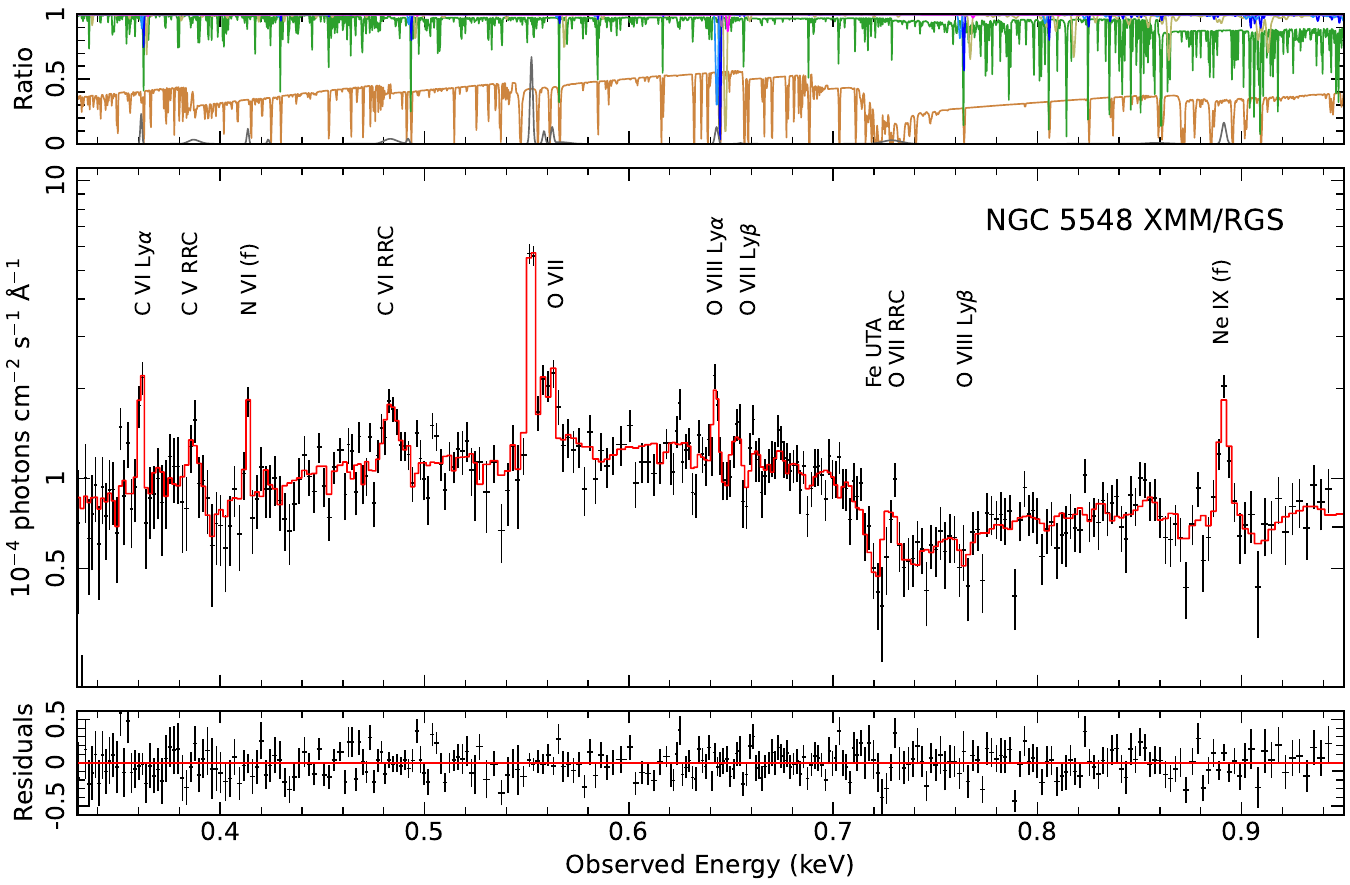}}
\vspace{-0.2cm}

\caption{XRISM/Resolve (top) and XMM/RGS (bottom) spectra of \ngc with the best-fit model (red; Tables~\ref{table_cont} and \ref{table_para}). Panels show close-ups of the strongest emission and absorption features. Spectra are binned for clarity. Fit residuals are defined as (data $-$ model)/model. Individual outflow components and their labels (Table~\ref{table_para}) are shown in the top panels, with the total emission line model overplotted in gray for reference.
\label{fig_spec}}
\end{figure*}

We jointly modeled the time-averaged \xrism/Resolve and \xmm/RGS spectra (Figs.~\ref{fig_spec} and \ref{fig_full}) using \spex\ {\tt v3.08.02} \citep{Kaas96,Kaas25} and its most up-to-date atomic database ({\tt {var calc new}} in \spex).
To avoid oversampling and to ensure statistically appropriate binning, we applied the \spex\ {\tt rbin} command to both the spectra and their response matrices. 
The underlying methodology of {\tt rbin}, which accounts for source statistics as well as the instrumental spectral resolution, is described in the \spex\ manual (see also \citealt{Kaas16}).
The contribution of the NXB to the Resolve spectrum (described in Sect. \ref{sect_resolve}) was accounted for using a {\tt file} model in \spex.
The Resolve spectrum was fitted over the 1.9--12~keV energy band, and the RGS spectrum over 0.3--1.9~keV, using C-statistics.
The Resolve and RGS spectra were assigned to different ``sectors'' in \spex, allowing for independent continua for each spectrum to account for continuum and obscuration variability between the two epochs, while all other components, including the ionized outflows, are linked in the joint modeling.
As described later in Sects.~\ref{sect_cont} and \ref{sect_abs}, the continuum and obscuration parameters vary between the two epochs, while the ionized outflow components remain consistent.

The cosmological redshift ({\tt reds}) was fixed at 0.017175 \citep{deVa91}, which in \spex corresponds to a luminosity distance of 74.5~Mpc assuming ${H_{0}=70\ \mathrm{km\ s^{-1}\ Mpc^{-1}}}$, $\Omega_{\Lambda}=0.70$, and $\Omega_{\rm m}=0.30$. Galactic absorption was modeled using the \spex\ {\tt hot} component \citep{dePl04,Stee05}, with the temperature fixed at its minimum possible value (0.001~eV) and the column density set to $\NH = 1.45 \times 10^{20}$~cm$^{-2}$ \citep{Wakk11}. Elemental abundances for all model components were fixed to the protosolar values of \citet{Lod09}.

\subsection{Broadband Continuum and Photoionization Modeling}
\label{sect_cont}

The intrinsic continuum model for the spectral energy distribution (SED) of \ngc\ was originally established in \citet{Meh15a} based on an extensive multi-wavelength campaign in 2013, and has since been adopted in subsequent studies, most recently \citet{Mehd24}. 
In this framework, the intrinsic continuum (prior to any absorption) consists of a {\tt comt} component that models the optical/UV disk continuum and the `soft X-ray excess' via warm Comptonization, together with a {\tt pow} component representing the hard X-ray power-law continuum.
We adopt this same broadband continuum model in our analysis of the 2025 Resolve and RGS spectra.
For the 2025 data, the normalization and photon index of {\tt pow} are fitted, while its low- and high-energy exponential cutoffs are fixed at 1~Ryd and 400~keV, respectively \citep{Meh15a}.
The normalization of {\tt comt} is also scaled to match the UV continuum level observed with HST and \xmm/OM. All other {\tt comt} parameters are fixed to the values derived from the 2013 campaign (see Table 2 in \citealt{Meh15a}): temperature of the seed photons $T_{\rm seed} = 0.8$~eV, temperature of the warm corona $T_{\rm e} = 0.17$~keV, and optical depth ${\tau = 21.1}$. These parameters already provide a continuum shape consistent with the current data and, in addition, cannot be independently constrained here: the \xrism/Resolve spectrum lacks sensitivity to the soft excess due to the closed gate valve, and the soft band in the \xmm/RGS data is affected by obscuration. As a result, refitting these parameters would not be required. The best-fit parameters of the {\tt pow} and {\tt comt} components for the 2025 \xrism\ and \xmm\ observations are listed in Table~\ref{table_cont}.

Photoionization calculations were performed using the \pion\ model \citep{Mill15,Meh16b} in \spex. 
We used the best-fit broadband continuum models (Table \ref{table_cont}) described above as the ionizing continua for photoionization modeling.
As detailed in \citet{Meh16b}, \pion\ normally carries out two steps: (1) computing the photoionization balance for the specified SED, and (2) producing the corresponding spectrum.  
We computed step (1) using the {\tt xabsinput} tool, which runs \pion\ to generate tables of ionic concentrations as a function of ionization parameter ($\xi$). 
These tables are then supplied to the \xabs\ model \citep{Stee03} within \spex, which performs step (2) by calculating the model spectrum.
This strategy yields absorption spectra that are fully consistent with \pion\ while avoiding repeated recalculation of the photoionization balance during fitting. 
As a result, the spectral fitting becomes significantly faster.
In Sect. \ref{sect_abs} we describe the \xabs fitting of the absorption in the Resolve and RGS spectra.

%

%

\subsection{Modeling of the X-ray Emission Features}
\label{sect_emit}

As the primary aim of this work is to characterize the absorption associated with the outflows, we adopt a simple but flexible prescription for the emission features. 
The goal is to obtain a well-fitted model of the line emission and underlying continuum, ensuring accurate measurement of the parameters of the absorption lines.
Emission lines are modeled using a delta function (\delt\ in \spex) convolved with a Gaussian profile (\vgau). 
Line centroids are fixed at zero velocity shift in the AGN rest frame, and for doublets we impose the theoretical line ratios. 
The Gaussian velocity widths of the emission lines are tied together in a physically consistent manner. 
Radiative recombination continua (RRCs) are modeled using the {\tt rrc} model in \spex, which includes Bremsstrahlung, two photon emission and free-bound radiation.
We account for the following emission features: \FeKa, \FeKb, Ni~K$\alpha$, and the H-like and He-like transitions of C, N, O, Ne, Si, and Fe. 
The velocity broadening, $\sigv$, of the H-like and He-like lines and RRCs is measured to be about 350~\kms in both the Resolve and RGS spectra.  
For the \fexxvi emission, an additional broad component is detected, with $\sigv \sim 3700$~\kms.

The \fexxv and \fexxvi features in the Resolve spectrum show hints of P-Cygni profiles, consistent with a photoionized wind that both absorbs and emits.
A detailed photoionization analysis of the emission lines will be presented in future work; for previous photoionization modeling of emission features in the archival RGS spectrum of \ngc, see \citet{Mao18}.
In our modeling most emission features are well described by a single velocity-broadening component. However, fitting the \FeKa\ complex requires three distinct {\tt vgau} velocity components with $\sigma_v \sim 240$, 1000, and 4200~\kms\ in order to reproduce its full profile.
Our fits also indicate the presence of an extremely broad emission component ($\sigma_v \sim 25000$~\kms), which may originate from the accretion disk. 
A detailed physical interpretation of this feature is beyond the scope of the present work and will be explored in a future study.
The adopted components together provide an excellent description of the continuum and emission-line features, enabling a robust modeling of the absorption features, as described in Sect.~\ref{sect_abs}.
Therefore, as long as the emission lines are well fitted (as done here), the spectroscopy and decomposition of the absorption lines (our main objective) remain robust with respect to the choice of physical models for the emission lines.

\subsection{Modeling of X-ray Absorption by AGN Outflows}
\label{sect_abs}

The {\tt pion}/{\tt xabs} framework described in Sect.~\ref{sect_cont} is used to model the absorption produced by the ionized outflows in \ngc. 
The term ``ionized outflows'' refers to both the highly ionized outflows in the Fe~K band observed with Resolve and the lower-ionization outflows (also called warm absorbers) in the soft band seen with RGS. 
The Resolve and RGS spectra are jointly fitted using the same ionized outflow parameters, since the two bands primarily probe different ionization components and are well described by a single set of parameters, with no statistical requirement for further differentiation.
Individual {\xabs} components are introduced to reproduce the full set of absorption features in both spectra. 
The number of required components is driven by the diversity of ionic species present and by the fact that some ions appear at more than one velocity. 
To account for the observed ionization range, we require four distinct components with different ionization parameters ($\xi$). 
Three of these further split into two velocity components each.
We label the components in order of decreasing $\xi$, with those at the same $\xi$ sub-labeled in descending order of column density (\NH), as listed in Table~\ref{table_para}. 

For each {\xabs} component we fit $\xi$, \NH, outflow velocity (\vout), and broadening (turbulent) velocity (\sigv). 
All ionized outflow components are adopted to fully cover the source, as this already yields a good fit to all the absorption lines.
Where appropriate, similar turbulent velocities are linked to reduce the number of free parameters. 
The highest-ionization components (A1 and A2) are required to model the \fexxvi absorption, while \fexxv and \ion{Fe}{24} absorption are primarily produced by Comps. B1 and B2. 
Components C1 and C2 account for the lower-ionization Fe species (mainly \ion{Fe}{18}--\ion{Fe}{20}) and some H-like ions seen with RGS, such as \ion{O}{8}. 
Finally, Comp. D models the lowest-ionization gas observed with RGS, including the unresolved transition array (UTA) of Fe M-shell ions, as well as some He-like and Li-like ions, such as \ion{O}{7} and \ion{O}{6}.

%
\begin{deluxetable}{l | c c}
\tablecaption{Best-fit parameters of the intrinsic continuum components and the obscurer for the \xrism/Resolve and \xmm/RGS spectra of \ngc.
\label{table_cont}}
\tablewidth{0pt}
\setlength{\tabcolsep}{11pt}
\tablehead{
Parameter & \colhead{\xrism} & \colhead{\xmm} \\
                                &   (July 2025) & (February 2025)
}

\startdata
Continuum: \\
\pow Norm.                      & $5.9 \pm 0.1$      &  $5.1 \pm 0.1$     \\
\pow $\Gamma$                   & $1.67 \pm 0.01$    &  $1.67$ (c)        \\
\comt Norm.                     & $6.0 \pm 0.1$      &  $6.0 \pm 0.1$     \\
\hline
Obscurer: \\
$C_f$                           & $ < 0.3$            &  $0.59 \pm 0.02$         \\
\enddata
\tablecomments{The normalization of the power-law component (\pow) is given in units of $10^{51}$ photons~s$^{-1}$~keV$^{-1}$ at 1~keV, while the normalization of the Comptonization component ({\tt comt}) is in $10^{55}$ photons~s$^{-1}$~keV$^{-1}$. The other parameters of {\comt} are fixed to $T_{\rm seed} = 0.8$~eV, $T_{\rm e} = 0.17$~keV, and $\tau = 21.1$. The ``(c)'' indicates that the parameter is coupled to another in our modeling.}
\end{deluxetable}

%
\begin{table}[!t]
\begin{minipage}[!t]{\hsize}
\setlength{\extrarowheight}{2pt}
\setlength{\tabcolsep}{3pt}
\caption{Best-fit parameters of the outflow components derived from the 2025 \xrism/Resolve and \xmm/RGS spectra of \ngc.
}
\label{table_para}
\centering
\small
\begin{tabular}{c | c c c c}
\hline \hline
Comp.   & \logxi             & \NH              & \vout           & \sigv           \\
        & (erg~cm~s$^{-1}$)  & ($10^{21}$~\cm)  & (\kms)          & (\kms)          \\
\hline
A1      &  $3.4 \pm 0.2$     & $30 \pm 10$      & $2730 \pm 60$   & $70 \pm 30$     \\
A2      &  ''                & $16 \pm 1$       & $1250 \pm 80$   & $70$ (c)        \\
B1      &  $2.72 \pm 0.02$   & $7 \pm 1$        & $980 \pm 20$    & $70$ (c)        \\
B2      &  ''                & $5 \pm 1$        & $240 \pm 70$    & $190 \pm 60$    \\
C1      &  $2.24 \pm 0.02$   & $8 \pm 1$        & $870 \pm 40$    & $<10$           \\
C2      &  ''                & $0.5 \pm 0.1$    & $2190 \pm 300$  & $190$ (c)       \\
D       &  $0.89 \pm 0.04$   & $7.2 \pm 0.3$    & $1030 \pm 40$   & $60 \pm 10$     \\                                                                                            
\hline
\multicolumn{5}{c}{C-stat} \\
\multicolumn{5}{c}{3039 / 3072 (Resolve) and 759 / 635 (RGS)} \\
\hline
\end{tabular}
\end{minipage}
\tablecomments{
Components are labeled with a letter in descending order of ionization parameter ($\xi$), with those sharing the same $\xi$ further sub-labeled by a number in descending order of column density (\NH). The symbol ``(c)'' indicates that the parameters are coupled.
}
\vspace{0.0cm}
\end{table}

%
In addition to absorption by the ionized outflows, \ngc is also absorbed by a partially covering obscurer \citep{Kaas14,Mehd22c,Mehd24}. 
We account for this component in our modeling of the 2025 spectra. 
As noted in \citet{Mehd22c}, the obscurer has gradually declined over the past decade, but it remains present and variable. 
To model the obscurer, we adopt the two-component \xabs obscurer model of \citet{Kaas14} derived from the 2013 multiwavelength campaign, in which denser (colder) clumps are embedded within a more diffuse (warmer) medium. 
As found in \citet{Mehd24}, the obscurer does not affect $\xi$, and therefore the absorption, of the other outflow components in the X-ray band.
Follow-up studies have shown that the covering fraction (\cf) of the warm phase is the primary parameter that varies \citep{Mehd16,Mehd22c}, and the long-term decline in obscuration is mainly due to changes in this value \citep{Mehd22c,Mehd24}. 
Accordingly, we allow \cf to vary when fitting the 2025 \xrism and \xmm spectra, while all other parameters of the obscurer are fixed to those of \citet{Kaas14}, as they do not require re-fitting.
The warm absorber components are illuminated by the partially obscured SED, where the intrinsic continuum is attenuated by the obscurer located interior to the warm absorber. Therefore, the warm absorber models for the two epochs were generated using different input SEDs (and hence different \xabs input files) that account for changes in the obscurer and intrinsic continuum variability.
The best-fit \cf values for both epochs are listed in Table~\ref{table_cont}.
The February 2025 \xmm observation suggests stronger obscuration (${\cf \sim 0.59}$) than the July 2025 \xrism observation ($\cf < 0.3$).
Such changes in $\cf$ are due to the time-variable nature of the obscurer on timescales of months, as shown by previous \swift monitoring studies of \ngc \citep{Mehd16,Mehd22c}.
We note that changes in the obscurer between the two epochs do not impact our modeling of the ionized outflows, as the X-ray outflows are unaffected by the presence of the obscurer \citep{Mehd24}, and also the spectral decomposition of the highly ionized outflows in the Fe~K band with Resolve is largely insensitive to obscuration.

%
\begin{figure}[!t]
\centering
\resizebox{\hsize}{!}{\includegraphics[angle=0]{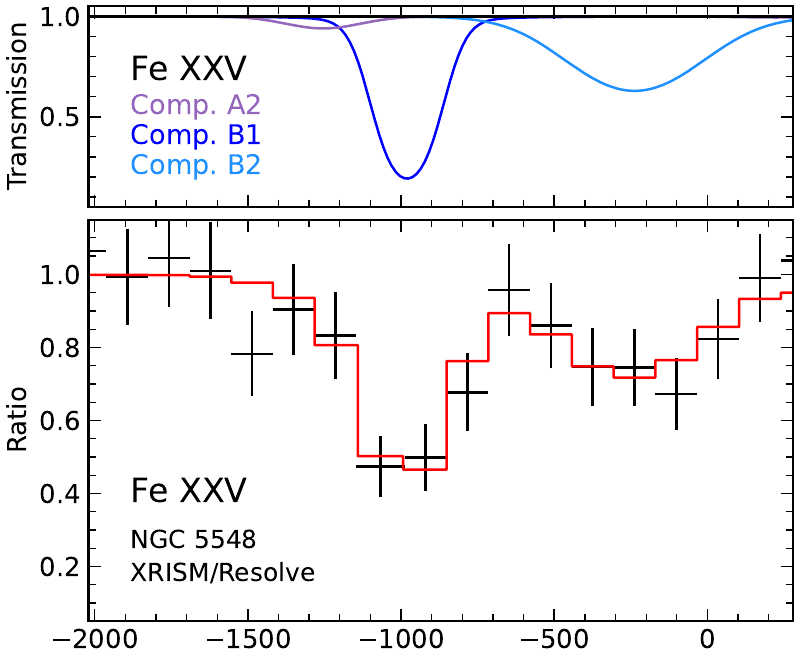}}\vspace{-0.0cm}
\resizebox{\hsize}{!}{\includegraphics[angle=0]{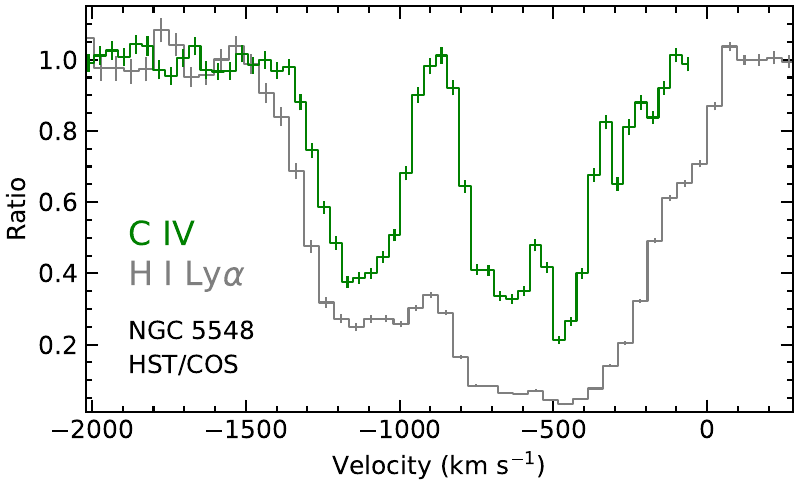}}
\caption{Absorption profile of the \fexxv resonance line in the \xrism/Resolve spectrum compared to \lya and \civ in the 2025 HST COS spectrum. The data are normalized to the continuum, showing the flux ratio on the y-axis. The red model in the middle panel corresponds to the best-fit model shown in Fig.~\ref{fig_spec}, with parameters listed in Table~\ref{table_para}. The top panel illustrates the contribution of individual model components to the \fexxv absorption. Negative velocities correspond to outflow, and zero velocity is defined in the \ngc rest frame at the wavelengths of the lines: \fexxv 1.8504~\AA\ (6.7004~keV), \civ 1548.19~\AA, and \ion{H}{1} \lya 1215.67~\AA.
\label{fig_profile}}
\end{figure}

\begin{figure}[!t]
\centering
\resizebox{\hsize}{!}{\includegraphics[angle=0]{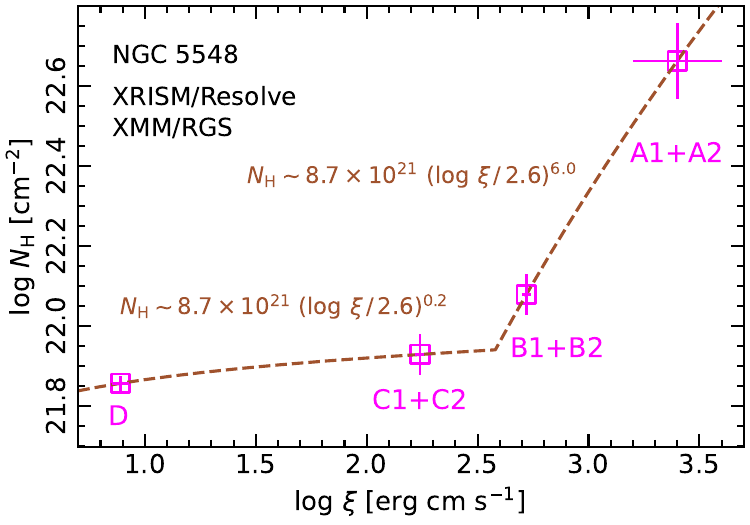}}\vspace{0.1cm}
\resizebox{\hsize}{!}{\includegraphics[angle=0]{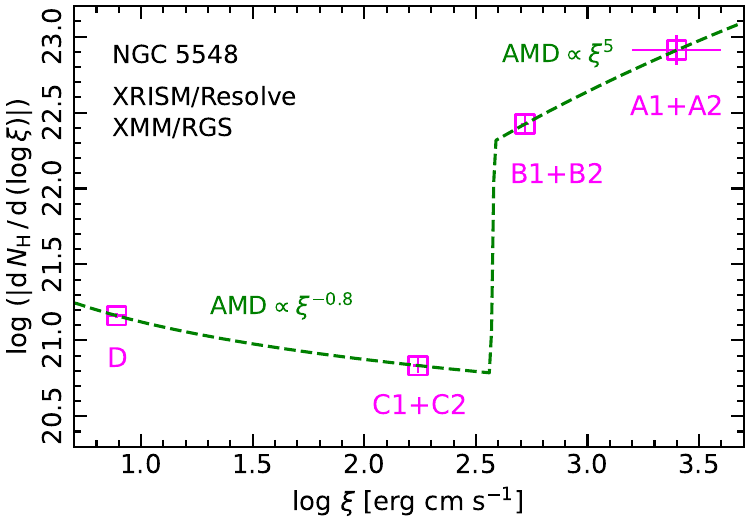}}\vspace{0.1cm}
\resizebox{\hsize}{!}{\includegraphics[angle=0]{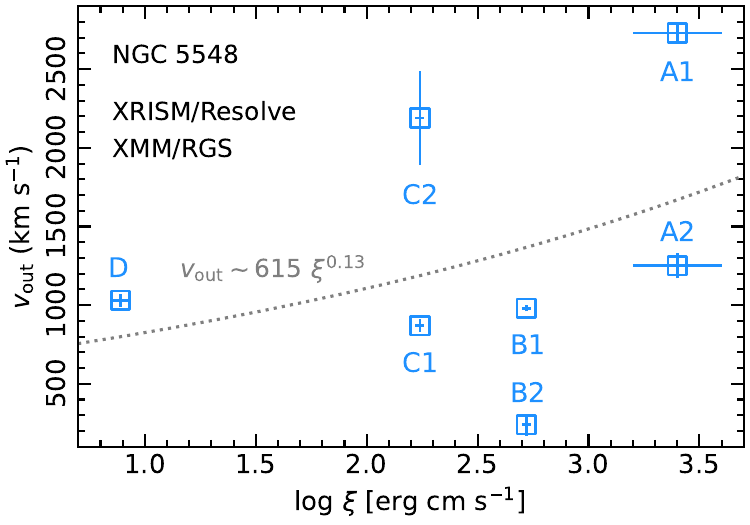}}
\vspace{-0.1cm}
\caption{Relations among the ionized outflow parameters (Table~\ref{table_para}) derived from the 2025 \xrism/Resolve and \xmm/RGS spectra of \ngc. The top panel shows the total column density (\NH) as a function of ionization parameter ($\xi$). The middle panel presents the absorption measure distribution (AMD), fitted with a broken power law, showing two distinct slopes above and below ${\log\xi \sim 2.6}$. The bottom panel shows the outflow velocity (\vout) versus $\xi$. Best-fit functions for each panel are indicated in the insets.}
\label{fig_rel}
\end{figure}

%
An overview of the \xrism/Resolve and \xmm/RGS spectra and our best-fit model is presented in Fig.~\ref{fig_full}.
A close-up of the spectra and best-fit model, highlighting the key absorption and emission regions, is shown in Fig.~\ref{fig_spec}.
The corresponding best-fit parameters of the ionized outflows are listed in Table~\ref{table_para}.
The reported statistical uncertainties on the fitted parameters correspond to the 1$\sigma$ confidence level.
The top panels of Fig.~\ref{fig_spec} also show the individual contributions of each {\tt xabs} component to the overall absorption, along with the underlying emission-line model. 
A close-up of the absorption profile of the \fexxv resonance line ($1s$--$2p$ transition), including the best-fit model and its individual {\tt xabs} components, is presented in Fig.~\ref{fig_profile}. 
For comparison, the absorption profiles of the \lya\ and \civ\ lines from the 2025 HST/COS spectrum are shown in the bottom panel of Fig.~\ref{fig_profile}. 
We note that the spectra in Figs.~\ref{fig_full} and \ref{fig_spec} are shown in the observed frame, whereas the profiles in Fig.~\ref{fig_profile} are in the rest frame of \ngc.
Finally, Fig.~\ref{fig_rel} illustrates the relationships between the key parameters of the outflow components. 
These results are discussed in detail in the following section.

%

\section{Discussion}
\label{sect_discuss}

\subsection{Ionization and Kinematic Structure of the Outflows}
\label{sect_structure}
%
%
The \xrism/Resolve spectrum of \ngc provides an unprecedented view of the highly ionized phase of its outflows.
For the first time in this source, the \fexxv\ and \fexxvi\ absorption lines are resolved, probing the most highly ionized regime of the wind. 
Combined with the \xmm/RGS spectrum, the data reveal a multiphase absorber spanning more than two orders of magnitude in the ionization parameter.
Our spectral modeling identifies four ionization components, three of which show two distinct velocity components. The outflow velocities cover a broad range from $\sim240$ to $\sim2730$~\kms.
The highest-ionization component, traced by \fexxv\ and \fexxvi, has $\log\xi \sim 3.4$, revealing a previously unseen phase of the wind in \ngc.
The observed outflow velocities are qualitatively similar to those previously reported in \ngc \citep{Stee05,Kaas14,Arav14}, as well as in other AGN such as NGC~3783 \citep{Kasp02,Mehd24}.
We do not detect any significant Fe~K absorption features due to UFOs in the time-averaged Resolve spectrum of \ngc, unlike those observed in NGC~3783 \citep{Mehd25,Gu25}.
Our results show that the highest-ionization component carries the largest column density, while progressively lower-ionization components contribute smaller columns. 
The increase of column density with ionization parameter is consistent with patterns seen in previous X-ray studies of AGN (e.g. NGC~4051; \citealt{Ogor22}; and the sample studies of \citealt{Beh09,Wang22}).
A similar trend is also evident from UV spectra alone (e.g. \citealt{Arav20}), where multiple ionization zones are present, with the high-ionization zone (comparable to typical X-ray warm absorbers) dominating the column density.
Our results in Fig.~\ref{fig_rel} show interesting trends, which we further discuss and interpret in Sect.~\ref{sect_origin}.

In this paper, we only examine the time-averaged 2025 spectra.
Time-resolved spectroscopy and variability studies of the outflows, which can provide useful constraints on their density and hence energetics (e.g., \citealt{Roga22}), will be investigated in a follow-up paper.
Nonetheless, we provide rough estimates of the kinetic luminosity of the outflows, under certain assumptions, such as the geometry and filling factor of the wind.
Kinetic luminosity ${L_{\rm kin} = 1/2 \, \dot{M}_{\rm out} \, v_{\rm out}^2 = 1/2 \, \mu \, m_{\rm p} \, N_{\rm H} \, R \, \Omega \, C_{\rm V} \, v_{\rm out}^3}$, 
where $\dot{M}_{\rm out}$ is the mass outflow rate, $\mu$ the mean atomic weight per proton ($\approx 1.43$, from \spex), $m_{\rm p}$ the proton mass, $\Omega$ the solid angle, $C_{\rm V}$ the volume filling factor, and $R$ the radial distance from the source. 
Following \citet{Blu05}, assuming the escaping wind velocity condition, we can place a constraint on the minimum $R$, and using the thin-shell scenario, we can place a constraint on the maximum $R$.
For these calculations, we use a SMBH mass ${M_{\rm BH} = 7 \times 10^7}$~$M_{\odot}$ \citep{Horn21}, the 1--1000 Ryd ionizing luminosity ${L_{\rm ion} = 1.2 \times 10^{44}}$~erg~s$^{-1}$ (from the broadband continuum modeling described in Sect.~\ref{sect_cont}), and a bolometric luminosity of ${L_{\rm bol} = 2.8 \times 10^{44}}$~erg~s$^{-1}$ (\citealt{Mehd24} and also Sect.~\ref{sect_cont}), corresponding to an Eddington ratio of ${L_{\rm bol} / L_{\rm Edd} = 0.03}$. 
We find that $R$ ranges from a minimum of 0.08~pc (Comp. A1) to a maximum of 139~pc (Comp. D). 
The density $n_{\rm H}$ ranges from a minimum of 10~${{\rm cm}^{-3}}$ (Comp. C2) to ${10^{5.8}~{\rm cm}^{-3}}$ (Comp. A1).
Using fiducial values of ${\Omega \sim 2\pi}$ (e.g. \citealt{Crens12}) and ${C_{\rm V} \sim 0.2}$ (e.g. \citealt{XRISM25}), we find that the total $L_{\rm kin} / L_{\rm bol}$ for all components in Table~\ref{table_para} is ${0.001 < L_{\rm kin} / L_{\rm bol} < 0.03}$.
For the highest ionization components (A1 + A2), the total ${L_{\rm kin} / L_{\rm bol} \approx 0.001}$.
For comparison, AGN feedback models require $L_{\rm kin} / L_{\rm bol} \gtrsim 0.005$ \citep{Hopk10} for a significant contribution to feedback.

The 2025 HST/COS spectrum of \ngc is most similar to that observed in 2011, albeit possibly with slightly less absorption. This epoch precedes the onset of the obscuring outflow, which now appears to be significantly diminished compared to a decade ago \citep{Mehd22c}.
The comparison of absorption line profiles for \fexxv, \civ, and \lya in Fig.~\ref{fig_profile} provides valuable insights and offers a model-independent way to examine their shapes. 
The lines exhibit both shared and distinct features, yet all display similar absorption over a broad velocity range, indicating a multi-component velocity structure in the outflowing gas.
The \fexxv profile clearly shows absorption troughs at two distinct velocities.
Almost all of the broadening seen in the \fexxv\ absorption (Fig.~\ref{fig_profile}) arises from the resonance line in each \xabs component, with the intercombination line contributing very little. 
Nevertheless, the spectral modeling includes all relevant transitions, enabling a robust determination of the intrinsic broadening for each component.
Figure~\ref{fig_profile} (top panel) shows how the components we have derived (Table \ref{table_para}) reproduce the observed profile of the \fexxv line.
While we attribute the intrinsic line broadening of each component to a single turbulence parameter, in reality each component may have an even more complex velocity structure, which would result in ``velocity shear'', which is challenging to constrain.
Nonetheless, by modeling the full \xrism/Resolve spectrum, the individual outflow components can be disentangled and parameterized, as shown in Figs.~\ref{fig_spec} and \ref{fig_profile} and Table~\ref{table_para}.

Table~\ref{table_para} shows that Comp. B2 has the lowest \vout ($\sim$240~\kms) and highest \sigv ($\sim$190~\kms) among all the components.
While a unique explanation for this is not possible, it may be because this component is not escaping and is falling back, thus producing an additional range of velocities along our line of sight. 
Alternatively, as this component is one of the weaker ones in terms of \NH, it may not be fully resolved into individual sub-components.
Furthermore, there appears to be an offset between the line centroids of the X-ray and UV absorption lines in Fig.~\ref{fig_profile}.
This may be an ionization effect, in which some regions of the wind are either too ionized to be seen in UV or not ionized enough to appear as \fexxv.
Additionally, this may be an acceleration effect on different phases of the wind.

The velocity broadening \sigv of some \xabs components cannot be accurately constrained and has therefore been tied between components in our modeling (Table \ref{table_para}).
This is primarily due to the limited S/N in the lines of these components, which prevents a reliable constraint on \sigv.
Additionally, the best-fit values for some components suggest low \sigv ($\sim 70~\kms$), which is below Resolve's instrumental resolution ($\sim 200~\kms$ at 6.7 keV).
Since the fit is driven by the full-band continuum and line modeling rather than the intrinsic line profile alone, the fitting procedure can favor artificially small \sigv values that marginally improve the overall C-statistic, without implying that the intrinsic line width is physically resolved.

The outflows in \ngc exhibit a wide range of ionization, from \fexxv to \civ, with each ion showing multiple velocity components.
Additionally, gas at similar velocities appears in multiple ions, which cannot originate from a single ionization phase.
These characteristics suggest a clumpy outflow structure, in which denser, cooler clouds are embedded within a more diffuse, hotter medium. 
This configuration is consistent with a broader multiphase circulation, in which hot, warm, and cooler phases are not isolated absorbers but instead exchange mass through condensation, entrainment, mixing, and recooling. 
Such behavior is naturally expected in the chaotic cold accretion (CCA) framework \citep{Gasp18,Macc21,Oliv22} and provides context for interpreting the partial kinematic overlap of the UV and X-ray absorption features in \ngc.

Clumpy outflows are commonly observed in joint X-ray and UV studies of Seyfert AGNs \citep{Mehd22c,Zaid24,Mehd25}, and have also been reported in near- or super-Eddington quasars such as PDS~456 \citep{XRISM25,Xu25} and PG~1211+143 \citep{Mizu26}.
This raises the question of whether clumpiness depends on luminosity or is more universal.
The \xrism results for \ngc and NGC~3783 \citep{Mehd25} suggest that clumpy structures may be a common feature of AGN outflows, potentially arising from thermal or hydrodynamic instabilities \citep{Take13,Dann20,Wate22}, rather than necessarily requiring extreme Eddington ratios.

\subsection{Physical Interpretation and Origin of the Outflows}
\label{sect_origin}
The absorption measure distribution (AMD), defined as ${{\rm AMD} \equiv |\mathrm{d}\,{N_{\rm H}}\, /\, \mathrm{d}\,(\log \xi)| \propto \xi^{a}}$ \citep{Beh09}, serves as a key diagnostic for probing the radial structure of AGN outflows \citep{Holc07,Beh09,Ster14,Adhi19,Kesh22}.
The slope $a$ of ${\log ({\rm AMD})}$ versus ${\log \xi}$ is a useful parameter, as it enables probing the hydrogen number density profile ${n_{\rm H} (r) \propto r^{-\alpha}}$ as a function of distance $r$ from the ionizing source. 
By measuring the slope $a$, the index $\alpha$ of the radial density profile can be obtained: ${\alpha = (1 + 2a) / (1 + a)}$ for a distance-driven AMD, or ${\alpha = 1 / (1 + a)}$ for a density-driven AMD \citep{Beh09}.
Thus, the radial density profile inferred from the AMD provides a link between observations and theoretical predictions for different wind launching and driving mechanisms.
Prior to the launch of \xrism, slopes of $\log \NH$ versus $\log \xi$ were found to range between 0.0 and 0.72 \citep{Kesh22}.
However, with \xrism, steeper slopes have also been observed (e.g., NGC~4151; \citealt{Xian25}), consistent with the magnetic wind models of \citet{Blan82}.

The relations between \NH and $\xi$ for the outflow components of \ngc (Fig.~\ref{fig_rel}, top panel) provide new insights into the nature of the AMD (Fig.~\ref{fig_rel}, middle panel) and the characteristics of the outflows in this AGN.
A single power-law or exponential function does not adequately describe the \NH-$\xi$ relation.
This is because, at higher ionization ($\log \xi > 2.6$), \NH rises sharply.
We have fitted a broken power-law function to the data to account for the different slopes, with a best-fit break at $\log \xi = 2.6$.
The fitted \NH versus $\log \xi$ functions are shown in the inset of Fig.~\ref{fig_rel} (top panel).
Using these fits, we computed the AMD shown in the middle panel of Fig.~\ref{fig_rel}.
The AMD exhibits two distinct regions with different slopes $a$: ${{\rm AMD} \propto \xi^{-0.8}}$ at ${\log \xi < 2.6}$ and ${{\rm AMD} \propto \xi^{5.0}}$ at ${\log \xi > 2.6}$.
This corresponds to radial density $n_{\rm H}$ profiles with very different indices $\alpha$ in the low- and high-ionization regimes.
Therefore, the observed AMD indicates a more complex radial density structure than a simple power-law profile, ${n_{\rm H} \propto r^{-\alpha}}$.
Previous studies have shown that the warm absorber in \ngc is a persistent feature with relatively stable physical properties over time \citep{Ebre16b}; thus, the overall AMD shape is not expected to be significantly affected by the use of non-contemporaneous observations, since any variability in \NH or $\xi$ of individual components is small compared with the broad range spanned by the AMD.

The \ngc AMD obtained from \xrism/Resolve reveals much finer structure, particularly at high $\xi$, than was accessible with earlier X-ray observatories.
This may explain why previous studies found the AMD across the X-ray band to be roughly consistent with a single power law \citep{Beh09,Wang22}, thus yielding different AMD slopes compared to ours.
Differences in AMD shapes between different AGN may reflect geometric effects, such as viewing angle, where the line of sight intersects different regions of the outflow or obscuring material. 
Also, in some cases, the soft X-ray band is strongly obscured (as in NGC~4151, \citealt{Xian25}), which can prevent the detection of lower-ionization components and thus alter the inferred AMD shape.

Previous AGN wind studies have explored correlations between \vout and $\xi$, which can provide insight into the density profile of the outflow (e.g. \citealt{Detm11}).
For example, a trend of ${\vout \propto \xi^{1/2}}$ would imply ${n_{\rm H}(r) \propto r^{-1}}$.
However, as shown in Fig.~\ref{fig_rel} (bottom panel), our results indicate that the \vout-$\xi$ distribution does not follow a clear trend.
Nonetheless, there is a general tendency for \vout to increase with $\xi$.
We have fitted a power-law function to the data, shown in the inset of Fig.~\ref{fig_rel} (bottom panel), which gives a rough approximation of $\vout \propto \xi^{0.13}$.
A more likely explanation for the wide scatter is that the outflows are driven by multiple mechanisms at different radii, so a simple analytic trend does not apply to all components.

The overall results of our investigation are consistent with a ``hybrid wind'' scenario, in which different components may have distinct origins and driving mechanisms.
The increasing AMD trends for highly ionized outflows are consistent with magnetically driven winds (e.g., \citealt{Fuku15}).
However, for the low-ionization ($\log\xi < 2.6$) ``warm absorber'' components, the declining shape of the AMD is likely influenced by multiple factors, including thermal driving \citep{Mizu19,Gang21}.
The decrease in the AMD at lower ionizations ($\log\xi < 2.6$) aligns with expectations for thermally driven winds \citep{Dyda17}.
This result in \ngc is similar to that found in the \xrism/Resolve study of NGC~3783 \citep{Mehd25}.
Additionally, the lack of a single analytic trend in the \vout-$\xi$ distribution points to a complex kinetic structure of the outflows, in which multiple factors may play a role, including radiation effects \citep{Prog00,Gius19,Wate21,Mizu21} and thermal instabilities inducing inhomogeneities (clumpiness) in the gas density.
A complex radial structure may also arise from turbulent mixing and interactions between distinct components of the outflows.
Thus, the ``hybrid'' wind scenario provides a plausible explanation for the observed pattern in the AMD.

Disk wind models (e.g. \citealt{Fuku17,Fuku24}) are capable of reproducing the spectral properties of diverse AGN outflows, ranging from UFOs to broad (obscurers) and narrow (warm absorber) absorption outflows.
Multiphase outflows with multiple ionization components and kinematic substructures can also be produced in chaotic cold accretion (CCA) models, where cold/warm clumps, mixing layers, and a hot phase coexist and exchange mass \citep{Gasp20}.
The partial similarity between the \fexxv and UV (\civ, \lya) absorption (Fig. \ref{fig_profile}) supports the idea that X-ray and UV absorbers represent different phases of the same multiphase flow cycle, involving condensation, shredding/mixing, and re-cooling, while differences in the line profiles reflect ionization stratification and phase-dependent covering.
The observed increase of $N_{\rm H}$ with the ionization parameter (Fig. \ref{fig_rel}) can be understood within this framework: high-$\xi$ gas is more volume-filling with longer effective path lengths, whereas low-$\xi$ gas resides in compact, dense clumps, yielding higher integrated columns at high $\xi$ without contradicting the dense nature of the cool phase \citep{Gasp17b}.
Similarly, the trend of $v_{\rm out}$ increasing with $\xi$ is consistent with high-$\xi$ gas being launched from deeper regions and more directly coupled to the fast, hot wind, while lower-$\xi$ phases correspond to entrained cloudlets and mixing layers at larger radii \citep{Gasp17b,Gasp17}.

In this Paper~I of our series, we focused on the time-averaged \xrism/Resolve and \xmm/RGS spectra to study the ionization and kinematic structure of the outflows in \ngc.
In follow-up papers of this \xrism series, we will investigate additional properties of the outflows, including their dynamics and evolution, by probing variability on both short and long timescales, and by comparing the 2025 spectra with previous X-ray and UV observations of \ngc.
To obtain a more comprehensive picture of the outflows, we will perform joint modeling of the outflows detected in both the UV (with HST) and X-rays (\xrism and \xmm).

\section{Conclusions}

The 2025 \xrism observation of \ngc provides the first view of highly ionized outflows in this AGN, demonstrating the diagnostic power of the Resolve microcalorimeter.
The data resolve individual ionization and velocity components in the Fe~K band, primarily traced by \fexxv and \fexxvi absorption lines.
By combining the \xrism/Resolve spectrum with \xmm/RGS data, we probe the outflows over a broad range of ionization across the X-ray band.
Our spectral modeling reveals a multiphase outflow structure, composed of four ionization components spanning more than two orders of magnitude in the ionization parameter $\xi$.
Three of these components are further resolved into two distinct velocity sub-components, yielding a total of six kinematic components with outflow velocities ranging from 240~\kms\ to 2730~\kms.
The outflows in \ngc are observed from X-rays to the UV, with multiple velocity components in each ion and similar velocities across different ions, indicating a clumpy, multiphase outflow structure.
An offset is observed between the line centroids of the X-ray (\fexxv) and UV (\civ\ and \lya) absorption lines, which may arise from the effects of ionization or acceleration within the wind.

We find that the column density \NH within the wind generally increases with $\xi$, although not as a single continuous function, and that higher-ionization components tend to exhibit larger outflow velocities.
The absorption measure distribution (AMD) is consistent with a broken power law, with ${{\rm AMD} \propto \xi^{-0.8}}$ at ${\logxi < 2.6}$ and ${{\rm AMD} \propto \xi^{5.0}}$ at ${\logxi > 2.6}$, indicating distinct regimes in the outflow properties.
These results point to a complex ionization and velocity structure in \ngc, consistent with a ``hybrid wind'' scenario in which different driving mechanisms and/or origins dominate in the low- and high-ionization regimes.
The combined \xrism, \xmm, and HST view provides a comprehensive picture of the outflow structure, yielding new insights into the origin and physical properties of AGN winds and their interaction with the circumnuclear environment.

\begin{acknowledgments}
M. Mehdipour acknowledges support from NASA XRISM grant 80NSSC25K0598. This work is also supported by NASA through a grant for HST program number 17591 from the Space Telescope Science Institute (STScI), which is operated by the Association of Universities for Research in Astronomy, Incorporated, under NASA contract NAS5-26555. SRON is supported financially by NWO, the Netherlands Organization for Scientific Research. The material is based upon work supported by NASA under award number 80GSFC24M0006. M. Dehghanian thanks NASA for the support provided by grants JWST-AR-06419, JWST-AR-06428, JWST GO5018, and JWST GO5354 from STScI. M. Gaspari acknowledges support from the ERC Consolidator Grant \textit{BlackHoleWeather} (101086804). We thank the anonymous referee for the constructive comments.
\end{acknowledgments}
\facilities{XRISM, XMM, HST(COS)}

\software{{\tt SPEX} \citep{Kaas96,Kaas25}}

\bibliographystyle{aasjournalv7}
\bibliography{references}{}

\end{document}